\documentclass[useAMS,usenatbib]{mn2e}

\usepackage{natbib}
\usepackage{graphicx}
\usepackage{dcolumn}
\usepackage{tikz}
\usepackage{pgf}
\usepackage{bm}
\usepackage{amsmath}
\usepackage{amssymb}
\usepackage{hyperref}
\usepackage{float}
\usepackage{booktabs}
\usepackage{tabularx}
\usepackage{soul}

\graphicspath{{figs/}}

\newcommand{\vect}[1]{\bm{#1}}

\newcommand{\Aq}[1]{\texttt{Aq-#1}}

\title[Shape of subhalos]{The shape of dark matter subhalos in the
Aquarius simulations}

\author[Vera-Ciro et al.]{
\parbox[t]{\textwidth}{
Carlos A. Vera-Ciro\textsuperscript{1,2}\thanks{E-mail: ciro@astro.wisc.edu}, 
Laura V. Sales\textsuperscript{3},
Amina Helmi\textsuperscript{2} and Julio F. Navarro\textsuperscript{4}
}\\ 
\\
\\
\textsuperscript{1} Department of Astronomy, University of Wisconsin, 2535
Sterling Hall, 475 N. Charter Street, Madison, WI 53076, USA\\
\textsuperscript{2} Kapteyn Astronomical Institute, University of
Groningen, PO Box 800, 9700 AV Groningen, The Netherlands \\
\textsuperscript{3} Max-Plank-Institut f\"ur Astrophysik,
Karl-Schwarzschild-Stra\ss e, 1, 85740 Garching bei M\"unchen, Germany
\\
\textsuperscript{4} Senior CIfAR Fellow. Department of Physics \& Astronomy,
University of Victoria, Victoria, BC, V8P 5C2, Canada \\
}

\begin{document}

\date{\today}

\pagerange{\pageref{firstpage}--\pageref{lastpage}} \pubyear{\the\year}

\maketitle
\label{firstpage}

\begin{abstract}

  We analyze the Aquarius simulations to characterize the shape of dark matter
  halos with peak circular velocity in the range $8<V_{\rm max}<200$ $\rm
  km/s$, and perform a convergence study using the various Aquarius resolution
  levels. For the converged objects, we determine the principal axis ($a \geq
  b \geq c$) of the normalized inertia tensor as a function of radius.  We
  find that the triaxiality of field halos is an increasing function of halo
  mass, so that the smallest halos in our sample are $\sim 40 - 50\%$ rounder
  than Milky Way-like objects at the radius where the circular velocity peaks,
  $r_{\rm max}$. We find that the distribution of subhalo axis ratios is
  consistent with that of field halos of comparable $V_{\rm max}$. Inner and
  outer contours within each object are well aligned, with the major axis
  preferentially pointing in the radial direction for subhalos closest to the
  center of their host halo. We also analyze the dynamical structure of
  subhalos likely to host luminous satellites comparable to the classical
  dwarf spheroidals in the Local Group. These halos have axis ratios that
  increase with radius, and which are mildly triaxial with $\langle b/a
  \rangle\sim0.75$ and $\langle c/a \rangle\sim0.60$ at $r \sim 1$ kpc. Their
  velocity ellipsoid become strongly tangentially biased in the outskirts as a
  consequence of tidal stripping.
\end{abstract}

\begin{keywords} methods: numerical - galaxies: dwarf - cosmology: dark
  matter.
\end{keywords}

\section{Introduction}
\label{sec:intro}

In the $\Lambda$ cold dark matter cosmological paradigm structures build
hierarchically, via the mergers of smaller objects \citep{Press1974, Gott1975,
  White1978, Blumenthal1984}. As mergers proceed, the innermost regions of
some of the progenitors survive, resulting in non-linear structures where a
wealth of substructure orbits the center of an otherwise monolithic dark
halo. Early $N$-body simulations showed that halos could host dozens of
substructures, down to masses near the numerical resolution limit
\citep{Tormen1997, Tormen1998, Ghigna1998, Klypin1999b, Klypin1999,
  Moore1999}. For systems like the Milky Way, current numerical simulations
have extended the dynamical range of resolved substructures by 4-5 orders of
magnitude \citep{Diemand2007, Diemand2008, Springel2008, Stadel2009}.

The properties of these substructures are of great interest since luminous
satellites, such as the population of dwarf spheroidal (dSph) galaxies in the
Local Group, are expected to be embedded in them \citep{Stoehr2002,
  Strigari2007, Boylan2012, Vera2012}. Furthermore, the large mass-to-light
ratios of dSph, which range from 10 s to 1000 s \citep{Mateo1998, Gilmore2007,
  Walker2012}, indicate that their internal dynamics is dominated by the dark
matter.  This suggests that the predictions of pure dark matter simulations
may be directly confronted with observations of these systems. For instance,
it has been suggested that they provide an optimal place to look for signals
of dark matter self-annihilation processes \citep{Kamionkowski2010} due to the
natural enhancement in density and the lack of significant contamination from
the baryonic component.

The availability of large samples of line-of-sight velocities for individual
stars in dSph galaxies offers new tests of the predictions of $\Lambda$ cold
dark matter ($\Lambda$CDM) regarding the structure of dark matter
subhalos. For instance, studies of $N$-body numerical simulations have shown
that the inner slope of the dark matter density profile is expected to be
cuspy in CDM models \citep{Navarro1996, Navarro1997}.  This seems to contrast
with the somewhat shallower slopes and even constant density cores proposed to
explain the motions of stars in local dwarf spheroidals \citep{Amorisco2011,
  Walker2011, Amorisco2012, Jardel2012}. This, however, is a subject of active
debate, since various authors have shown that the stellar kinematics of Milky
Way dwarfs are also consistent with the NFW cuspy profiles
\citep{Battaglia2008, Walker2009, Strigari2010, Breddels2012,Breddels2013}.

Most of the dynamical modeling performed in the studies of Local Group dwarf
spheroidals relies on simple assumptions about the structure of their dark
matter component.  In particular, spherical symmetry and specific anisotropy
profiles have been extensively assumed.  The orbital anisotropy has been taken
to be constant \citep{Richstone1986, Lokas2002, Lokas2005, Lokas2009,
  Walker2009}, or radially dependent \citep{Kleyna2001, Wilkinson2002,
  Battaglia2008, Strigari2008, Wolf2010, Amorisco2011}, while in Schwarzschild
modeling or in made-to-measure $N$-body methods it does not need to be assumed
\citep{Long2010,Jardel2012, Breddels2012}. The selection of geometric shape
for the dark matter potential can also be relaxed. For example,
\cite{Hayashi2012} considered axisymmetric dark matter halos to model some of
the Milky Way dSph galaxies.

For isolated galaxies, numerical experiments of $\Lambda$CDM have clear
predictions for these quantities. The shapes of (isolated) dark matter halos
in the mass range $10^{10}-10^{15}$ M$_{\odot}$, are generally found to be
triaxial, with axis ratios depending on the mass of the object
\citep{Frenk1988, Dubinski1991, Warren1992, Cole1996, Thomas1998, Jing2002,
  Bailin2005, Kasun2005, Hopkins2005, Allgood2006, Bett2007, Hayashi2007,
  Kuhlen2007, Stadel2009, Diemand2009, Munoz2011, Schneider2012}. In terms of
their internal kinematics, the velocity ellipsoid is close to isotropic near
the center of halos and becomes mildly radial towards the outskirts
\citep{Wojtak2005,Hansen2006, Ludlow2011}.

For subhalos, however, less is known because of the demanding numerical
resolution needed to model properly low mass halos orbiting within hosts of
Milky Way mass.  This situation has recently improved with simulations that
are now able to successfully sample the mass function on these scales, such as
the Via Lactea, CLUES, GHALO and Aquarius simulations \citep{Diemand2007,
  Diemand2008, Springel2008, Stadel2009, Libeskind2010}. For instance, using
the Via Lactea simulations \cite{Kuhlen2007} found that subhalos are also not
spherical, although the effect of tides tends to make subhalos rounder than
comparable objects in the field.  These results prompt questions about the
validity of some of the assumptions involved in the mass modeling of stellar
kinematics in dwarfs. And although the orbital anisotropy of the stars in a
dSph is likely unrelated to that of dark matter (and associated to the
formation history), it is nonetheless valuable to explore the dynamical
structure of subhalos because they provide the main potential.

A detailed study of the shape of the Milky Way mass Aquarius halos was
presented in \citet{Vera2011}. Here we extend this analysis to lower mass
objects, both subhalos of the main central halo and field halos, up to $1.5
h^{-1} \rm Mpc$ from the center of the main Aquarius halo.  The paper is
organized as follows. In Section \ref{sec:numerical-modelling} we describe the
numerical simulations, introduce the methods we use to determine halo shapes
and explore the convergence of the results.  In Section \ref{sec:shape-today}
we compare the properties of subhalos and isolated objects of similar mass.
We analyze subhalo shapes in the context of the kinematic modeling of dwarfs
around the Milky Way in Section \ref{sec:shape-mw} and summarize our main
results in Section \ref{sec:conclusions}.

\section{Shape measurements and convergence}
\label{sec:numerical-modelling}

We use the suite of cosmological $N$-body simulations from the Aquarius
project \citep{Springel2008}. These consist of six $\sim 10^{12}\;{\rm
  M}_\odot$ $\Lambda$CDM halos (\Aq{A} to \Aq{F}), re-simulated with five
different levels of resolution within the cosmology $\Omega_0 = 0.25$,
$\Omega_{\Lambda} = 0.75$, $H=100h \;{\rm km\;s^{-1}Mpc^{-1}}$, $h=0.75$.  The
simulations use the zoom-in technique, with a high-resolution region that
extends at $z=0$ up to $\sim 2$ $h^{-1}$Mpc from the center of each main
halo. This exceeds the typical virial radius of the Aquarius halos by 5--10
times and allows us to identify isolated halos that have been unaffected by
tidal forces \citep[see ][for further details]{Springel2008}.

In most of the analysis that follows we focus on the level-2 resolution runs,
with a mass per particle $m_p\approx 10^{4}\;{\rm M}_{\odot}$. However, we use
the other Aquarius levels to test the convergence of our results. Halos and
subhalos are identified using the {\sc subfind} algorithm
\citep{Springel2001}. We keep all structures identified with at least 20
particles. We will call the central subhalo of a group a \emph{field/isolated
  halo}. In this work we consider field halos up to a distance of $1.5h^{-1}$
Mpc from the center of the main Aquarius halo, to avoid contamination of
low-resolution particles.

To measure the shape of halos in the simulations we follow the same approach
as \citet{Vera2011} and iteratively compute the inertia tensor in ellipsoidal
regions. At a given radius, the algorithm begins with a spherical contour
which is reshaped and reoriented according to the principal axis of the
normalized inertia tensor for the encompassed material, until convergence is
reached \citep{Allgood2006}.  More specifically, we define the normalized
inertia tensor as
\begin{equation} I_{ij} = \sum_{\vect{x}_k \in V}
\frac{x_k^{(i)}x_k^{(j)}}{d_k^2},
\end{equation} where $d_k$ is a distance measure to the $k$-th particle and
$V$ is the set of particles of interest. Representing dark matter halos as
ellipsoids of axis lengths $a \ge b \ge c$, the axis ratios $q=b/a$ and
$s=c/a$ are the ratios of the square-roots of the eigenvalues of $\vect{I}$,
and the directions of the principal axes are given by the corresponding
eigenvectors. Initially, the set $V$ is given by all particles located inside a
sphere which is re-shaped iteratively using the eigenvalues of $\vect{I}$. The
distance measure used is $d_k^2 = x_k^2 + y_k^2/q^2 + z_k^2/s^2$, where $q$
and $s$ are updated in each iteration. In practice we find that the algorithm
converges (i.e. the variation in the shape between successive iterations is $<
1$\%) when there are as few as 200 particles in set $V$. Notice that this 
is a more stringent criteria than required by plain identification of bound
objects in {\sc subfind}, which is here taken to be only 20 particles.

In \citet{Vera2011} we showed that shapes can be robustly measured from the
\emph{convergence radius}, $r_{\rm conv}$ outwards
\citep{Power2003,Navarro2010}. $r_{\rm conv}$ is defined so that the ratio
between the local relaxation time and the dynamical time at the virial radius
equals $\kappa$ where:
\begin{equation} \kappa(r) = \frac{\sqrt{200}}{8} \frac{N(r)}{\ln N(r)} \left[
    \frac{\overline{\rho}(r)}{\rho_c}\right]^{-1/2},
\end{equation}

\noindent where $N(r)$ is the number of particles inside the radius $r$,
$\overline{\rho}$ is the spherically averaged density and $\rho_c$ the
critical density. We adopt the value $\kappa=7$ because this guarantees that
the circular velocity profiles of the main halos are accurate to better than
$2.5\%$ \citep{Navarro2010}. Note that this equation has to be numerically
solved for $r_{\rm conv}$ with fixed $\kappa$ for each object in the
simulation. As a rule-of-thumb, we find that requiring a minimum of $\sim
10000$ particles enclosed within the radius of interest (i.e. $r_{\rm max}$ or
$r_{95}$) ensures that more than $90\%$ of the subhalos satisfy this
constraint.

\begin{figure}
  \includegraphics[width=0.48\textwidth]{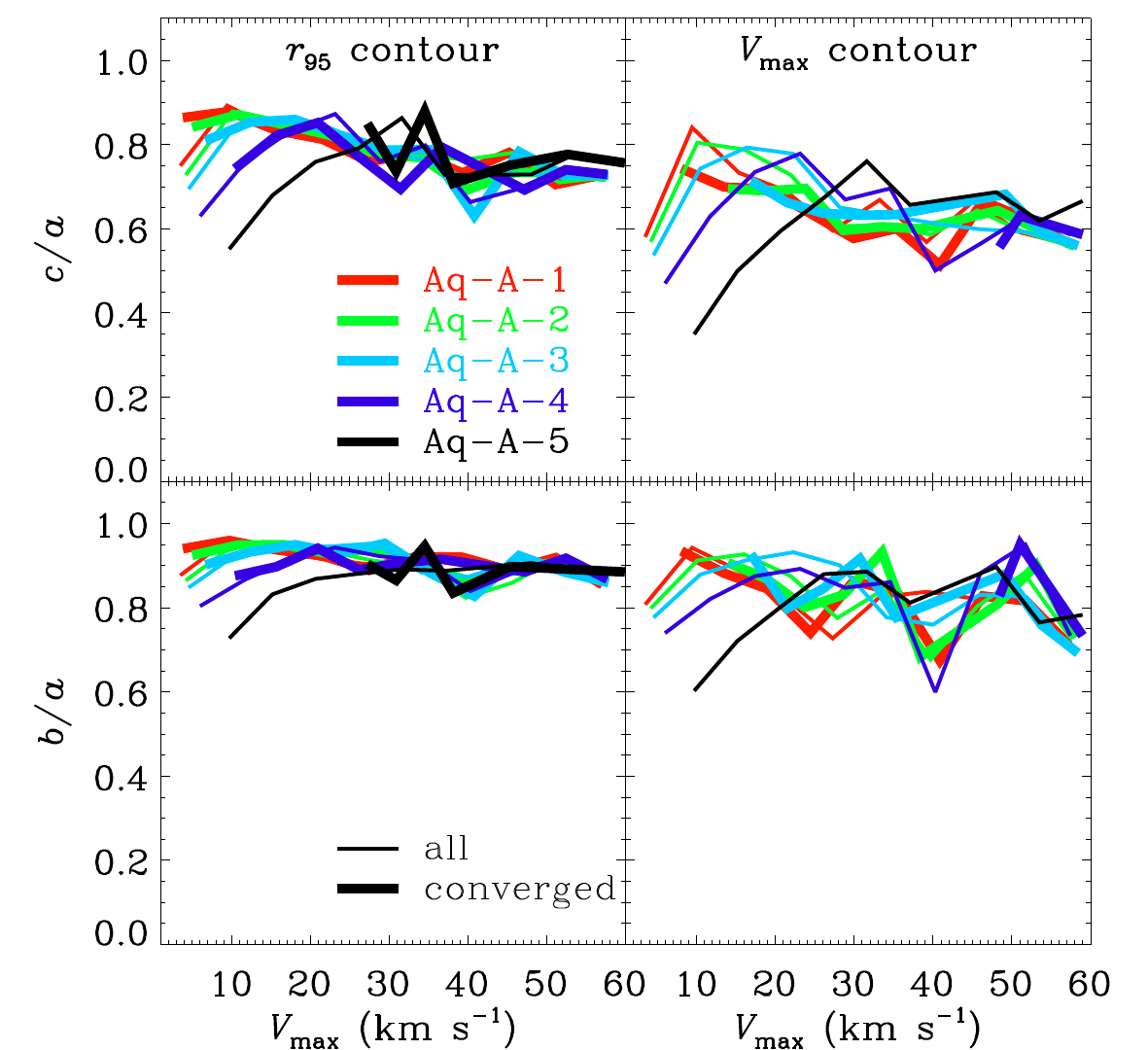}
  \caption{Shape of the $r_{95}$ contour (left) and $V_{\rm max}$ contour
    (right) as a function of $V_{\rm max}$ for subhalos of the main halo at
    five different resolutions. With thick lines we plot subhalos for which
    $r_{\rm conv} \le r_{95}$ (left) and $r_{\rm conv} \le r_{\rm max}$
    (right), where $r_{\rm conv}$ is such that $\kappa (r_{\rm conv}) = 7$.}
  \label{fig:convergence}
\end{figure}

Fig.~\ref{fig:convergence} shows that the same criteria applied to our sample
of subhalos also ensure convergence of halo shape estimates. We compare the
results for the \Aq{A} run in all resolution levels 1--5 (red to black,
respectively).  The left panels show, as a function of halo maximum circular
velocity $V_{\rm max}$, the mean axis ratios computed at $r_{95}$, here
defined as the ellipsoidal contour enclosing the $95\%$ most bound particles
identified by {\sc subfind}.  The thin lines correspond to the entire sample
of subhalos, whereas the thick curve shows only ``converged'' objects (those
where $r_{95} \geq r_{\rm conv}$). At level 2, the one used for most of our
analysis, the mean axis ratios agree with the highest resolution run \Aq{A-1}
to better than $5\%$ across the full spectrum of ``converged'' subhalos.

A similar conclusion is reached for the inner regions of subhalos, as shown by
the right panels of Fig.~\ref{fig:convergence}. Here, $c/a$ and $b/a$ are
computed at the radius of the peak circular velocity $V_{\rm max}$, which is
typically approximately nine times smaller than $r_{\rm vir}$ for field halos
and approximately six times smaller than $r_{95}$ for subhalos. The number of
objects for which $r_{\rm conv} < r_{\rm max}$ is roughly ten times lower than
those with $r_{\rm conv} < r_{95}$. This explains the relatively more noisy
behavior of the curves on the right column compared to those on the left
(especially for the lowest two resolution runs, where typically less than ten
objects satisfy the convergence condition).  In general, a subhalo whose shape
has converged at the $r_{95}$ radius has not necessarily converged at the
$r_{\rm max}$ radius.

Besides the mean trends shown in Fig.~\ref{fig:convergence}, we have also
explored the convergence of halo shapes on an object-by-object basis. In order
to do this, we identify the same subhalos in several resolution levels of the
\Aq{A} halo by matching the Lagrangian positions of all the particles assigned
to a substructure by {\sc subfind} back in the initial conditions \citep[see
Section 4.2][for further details]{Springel2008}. In addition to this
criterion, we impose a maximum deviation on the orbital path of matched
objects in different level runs. This is to ensure that the evolution of each
subhalo has been comparable in the different resolution runs also in the
non-linear regime.  More specifically, we define
\begin{equation} \Delta_{\vect{r}}^2 = \frac{1}{N}\sum_{\rm
snapshot}\frac{|\vect{r}_3(t) - \vect{r}_2(t)|^2}{|\vect{r}_2(t)|^2},
\end{equation} with $\vect{r}_j$ is the position\footnote{The positions of all
halos and subhalos are defined by the particle with the minimum potential
energy.} of the subhalo with respect to the main halo center at the $j$-th
resolution level. This is computed for every snapshot from the time the object
is first identified in the simulation box until present day. We consider only
structures for which $\Delta_{\vect{r}} \le 0.1$.

\begin{figure}
  \includegraphics[width=0.48\textwidth]{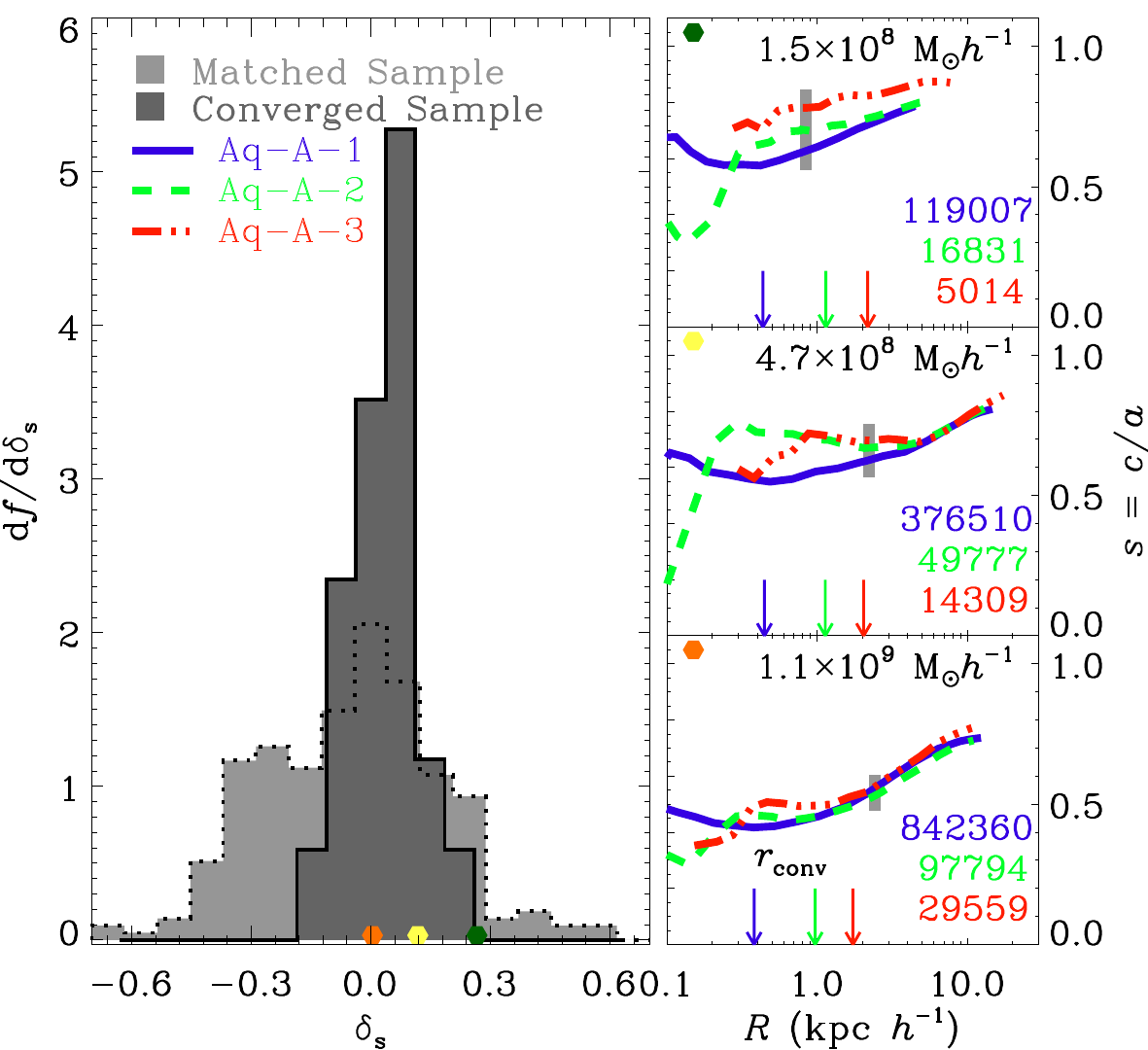}
  \caption{Left: distribution function of the deviation of the major-to-minor
    axis ratio with respect to the highest resolution simulation
    $\delta_s$. The light-gray histogram shows the distribution for the
    matched sample, while the dark-gray shows the results for the sample that
    satisfies that $r_{\rm conv}$ (vertical arrows in the right panels) is
    smaller than $r_{\rm max}$ (vertical gray line in the right panels) in all
    resolutions. The converged sample is narrower by a factor of 5. Right:
    axis ratios as function of position for three different subhalos with the
    quoted number of particles in the respective resolutions.}
  \label{fig:golden-sample}
\end{figure}

A total of 260 substructures are successfully matched in all levels 1,2 and 3
of the \Aq{A} halo by this procedure.  For each object, we define
$\delta_s=s_3/s_1-1$, where $s=c/a$ measured at $r_{\rm max}$ and the lower
indices indicate the resolution level (1 and 3 in the example above). By
construction, $\delta_s \sim 0$ for well-converged objects.  We show the
distribution of $\delta_s$ in Fig. ~\ref{fig:golden-sample}.  The light gray
histogram corresponds to all matched objects, and is significantly broader
than the distribution for the converged sample (i.e. the subset of the matched
sample formed only with the subhalos that have $r_{\rm max} > r_{\rm conv}$),
which is shown in dark gray.

We illustrate this more clearly on the right panels in
Fig.~\ref{fig:golden-sample}, which show the behavior of $c/a$ as a function
of radius $R = (abc)^{1/3}$ for three subhalos in the sample.  Small colored
dots indicate their $\delta_s$ value on the histogram on the left. The various
curves correspond to the results for different resolution levels as indicated
by the labels. For each subhalo the convergence radius $r_{\rm conv}$ is
marked with a vertical arrow and the position of the $V_{\rm max}$ contour by
a vertical thick gray line. The top panel shows a typical example of an
unconverged object: the peak of the circular velocity occurs at a smaller
radius than $r_{\rm conv}$ for levels 2 and 3. On the other hand, subhalos in
the middle and bottom panels have $r_{\rm max} > r_{\rm conv}$ and have
therefore converged (according to our criterion) at all these levels.

Notice that a large number of particles do not guarantee convergence. For
instance, the unconverged object on the top right panel of
Fig. \ref{fig:golden-sample} has $\sim 5000$ and $\sim 17000$ particles in
levels 3 and 2, respectively. These are significantly larger than the values
previously used in the literature \citep[e.g., ][]{Kuhlen2007, Knebe2008,
  Knebe2008b}, and highlights the need to impose a second criterion to measure
individual shapes reliably.  With our criterion, for only $\sim 2$\% of the
halos with $5000-10000$ particles have the shapes at the $V_{\rm max}$ contour
converged (i.e. $r_{\rm conv} > r_{\rm max}$).  The situation improves
significantly for the $r_{95}$ contour, where 99.6\% of such objects have
converged.

The distribution of axis ratios for converged objects shown in the left panel
of Fig. ~\ref{fig:golden-sample} has a standard deviation $\sigma=0.08$,
meaning that 68.3\% of the objects shapes determined at the Aquarius level-3
deviate less than 8\% from their value in the highest resolution run. Since we
focus on the level-2 runs for the analysis that follows, we expect resolution
effects in our sample to be negligible.

The above discussion shows that our criterion for convergence is relatively
strict.  There are 21403 subhalos with at least 200 particles within the
$r_{95}$ radius in all the Aquarius simulations, and we find that the inertia
tensor algorithm converges for 11483 subhalos at the $r_{\rm max}$ contour,
and for 13970 at the $r_{95}$ contour. If we now impose our convergence
criteria, there remain 412 and 6072 subhalos with well-determined shapes at
the $r_{\rm max}$ and $r_{95}$ contours, respectively.  For halos in the field
our convergence criteria leads to a reduction of 96\% and 35\% for the $r_{\rm
  max}$ and $r_{95}$ contours respectively. As expected, there is a larger
proportion of field objects whose shapes can be measured at the $r_{95}$
contour.  However, despite this significant reduction in sample size, we have
gained in the reliability of the shape determination for {\it individual}
halos.

Therefore, in the next section we focus on those halos which satisfy our
convergence criteria.

\section{Halo shapes as a function of mass and environment}
\label{sec:shape-today}

\begin{figure}
  \includegraphics[width=0.48\textwidth]{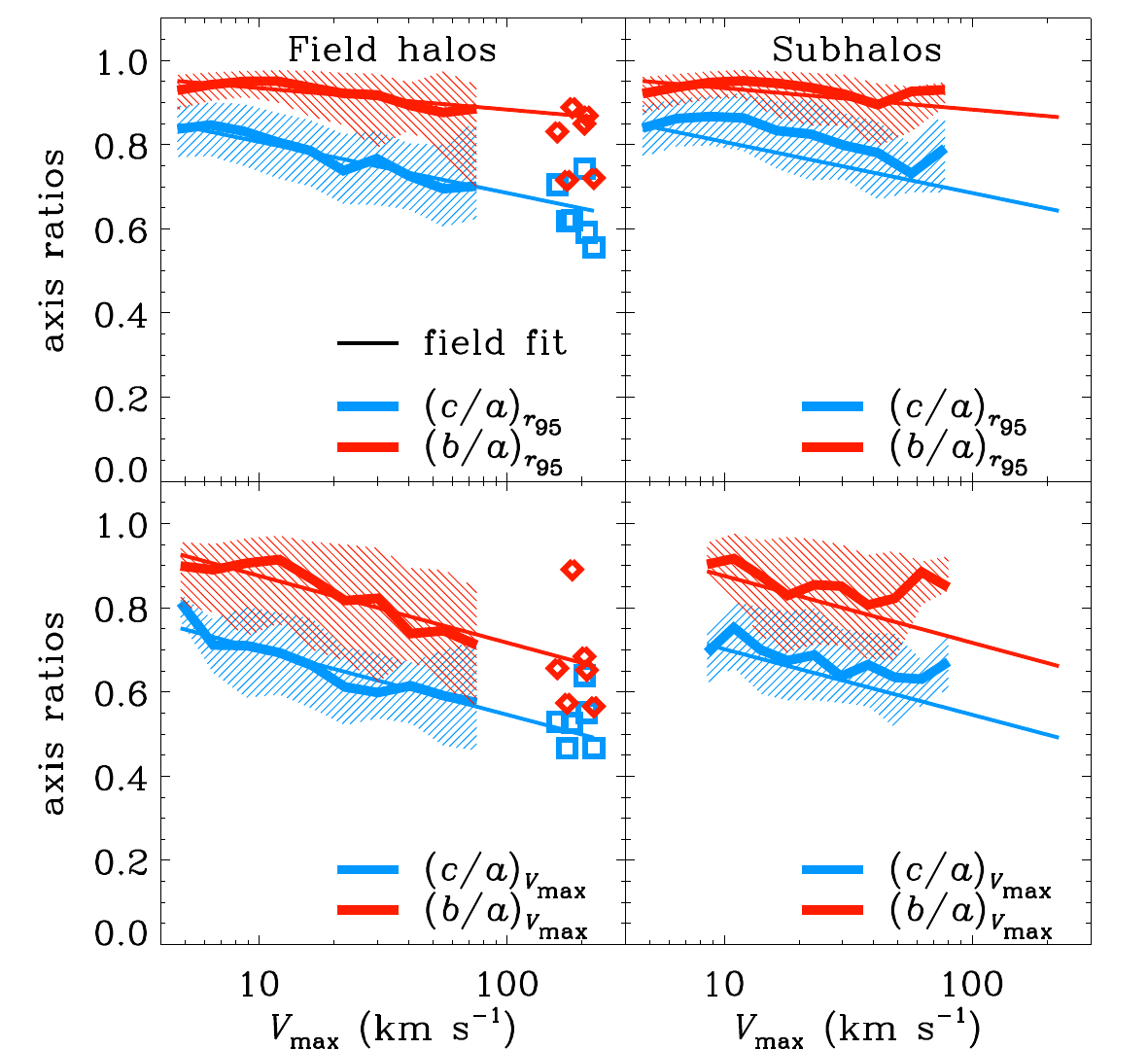}
  \caption{Shape as a function of $V_{\rm max}$ for field halos (left) and
subhalos of the main halos in the suite of Aquarius simulations (right). Thick
lines represent the median of the distribution of converged structures and the
shadowed region represents $\pm1\sigma$ equivalent dispersion around the
median. Thin lines are fits to the objects in the field. The diamonds and
squares indicate the axis ratios of the main Aquarius halos.}
  \label{fig:shape-vs-vmax}
\end{figure}

We proceed to characterize the variations in the axis ratios $b/a$ and $c/a$
of dark matter halos according to their mass or, equivalently, their maximum
circular velocity. The left column of Fig. ~\ref{fig:shape-vs-vmax} shows
$b/a$ and $c/a$ for isolated objects measured at the $r_{95}$ radius (labeled
\emph{$r_{95}$ contour}) and at $r_{\rm max}$ (\emph{$V_{\rm max}$ contour}) in
the top and bottom panels, respectively.  A thick line indicates the median
trend of our sample and the open symbols at the high mass end show the results
for the main Aquarius halos from \citet{Vera2011}.  In agreement with previous
work, we find that axis ratios tend to decrease gently with $V_{\rm max}$
\citep{Allgood2006, Maccio2007, Hahn2007, Bett2007, Munoz2011}, although we
now explore a different mass regime.

Inspection of the top and bottom panels shows that the dependence of the axis
ratios with circular velocity is somewhat steeper when measured at $r_{\rm
max}$ than at the $r_{95}$ contours. Typically, our lowest mass objects have
inner axes that are rounder by $40-50\%$ than those of Milky Way-like
halos. Nevertheless, the scatter from object to object at fixed circular
velocity also is larger at $r_{\rm max}$, as indicated by the shaded regions.

A comparison between the left and the right column of
Fig.~\ref{fig:shape-vs-vmax} reveals that there are only small differences
between subhalos and isolated objects. To make this comparison easier we
overplot in the panels on the right the linear fits obtained for field
halos. This shows that, on average, subhalos are slightly more spherical than
field halos at a given $V_{\rm max}$, but differences are well within the
scatter in the samples. The number of converged objects in the case of
subhalos is 385 and 1522 for $V_{\rm max}$ and $r_{95}$ contours, respectively.

\begin{figure}
  \includegraphics[width=0.45\textwidth]{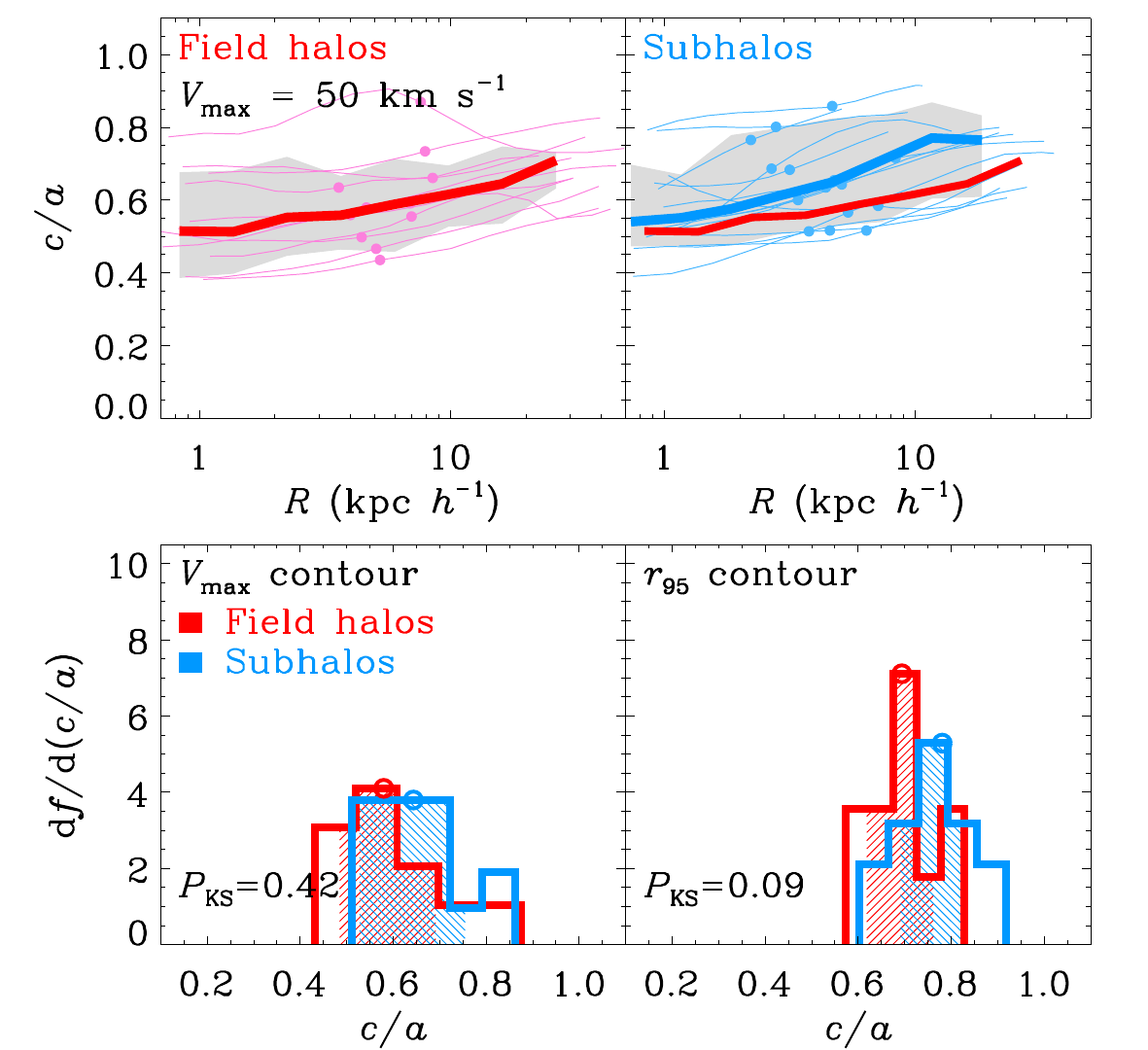}
  \caption{Top: the thin lines show the minor-to-major axis ratio profiles as
    a function of radius for objects with $V_{\rm max} \sim 50$ km s$^{-1}$.
    The thick lines correspond to the median and the shadowed region is
    $1\sigma$ equivalent scatter for the same sample of objects in the field
    (left) and subhalos (right). The median for field halos is also plotted in
    the right panel for comparison. Bottom: distribution of the axis ratios of
    the $V_{\rm max}$ (left) and $r_{95}$ (right) contour for the field halos
    (red) and subhalos (blue) plotted in the top panels. Although the median
    axis ratio (open circle) is slightly larger for subhalos the differences
    are well within the scatter.}
  \label{fig:shape-hist}
\end{figure}

Could the differences between field halos and subhalos be caused by measuring
shapes at different physical radii? It has been shown in the literature that
tidal evolution can significantly decrease $r_{\rm max}$ in satellites while
affecting $V_{\rm max}$ significantly less \citep{Hayashi2003, Kravtsov2004,
  Penarrubia2008}. In that case, the measurement of the halo shape at $r_{\rm
  max}$ would be at a smaller radius for a subhalo than for a halo in the
field with the same $V_{\rm max}$, and the same holds for the $r_{95}$
contour. We address this in Fig. \ref{fig:shape-hist}, where we show the
minor-to-major axis profiles for individual field halos (left) and subhalos
(right) of similar mass ($V_{\rm max} \sim 50$ km s$^{-1}$).  The solid dots
show the location of $r_{\rm max}$ for individual objects; they indicate that
the radii of the peak circular velocity are comparable in both samples and
therefore show that this can not be reason for the different trends reported
in Fig.~\ref{fig:shape-vs-vmax}. We thus confirm that, on average, subhalos of
a given $V_{\rm max}$ are slightly more spherical than comparable field halos
at all radii, particularly in the outskirts.  Kolmogorov--Smirnov tests
indicate that the difference between both samples is statistically significant
only at the $r_{95}$ contours (the Kolmogorov--Smirnov probability is 0.09 in
that case versus 0.42 at $r_{\rm max}$).  However, the differences are well
within the object-to-object scatter (see the bottom panel
Fig.~\ref{fig:shape-hist}).

\begin{figure}
  \includegraphics[width=0.48\textwidth]{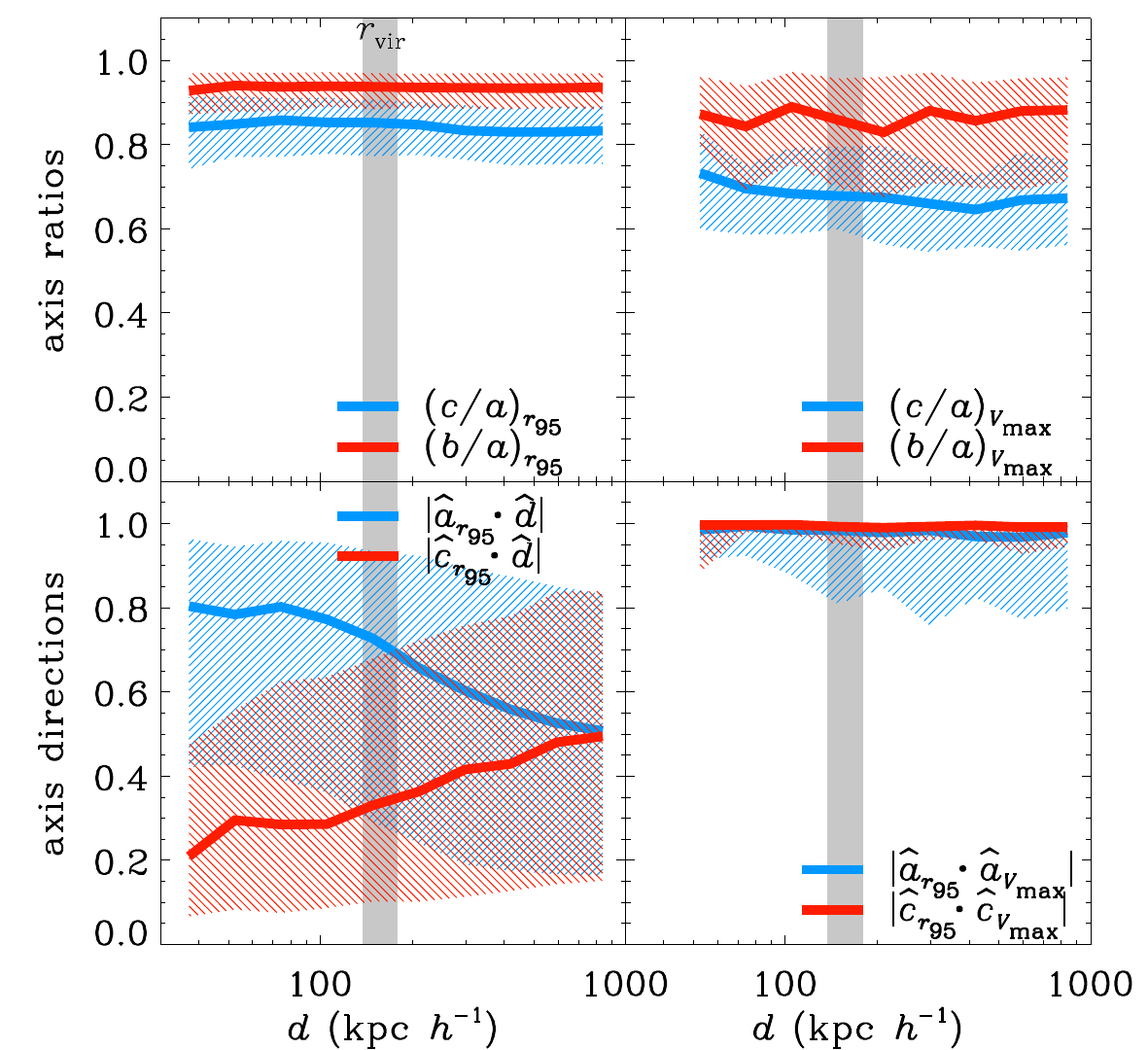}
  \caption{Shape and orientations of subhalos and objects in the field as a
    function of distance to the main halos $d$ in each of the Aquarius
    simulations. The gray line indicates the virial radii of the main halos. }
  \label{fig:shape-vs-d}
\end{figure}

The similarity between the subhalo and field populations apparent in
Figs~\ref{fig:shape-vs-vmax} and \ref{fig:shape-hist} explains the lack of
appreciable trends as a function of distance $d$ to the center of the main
Aquarius halos, shown in Fig.~\ref{fig:shape-vs-d}. The typical axis ratios
measured at the $r_{95}$ as well as $V_{\rm max}$ contours do not depend on
the distance to the host center up to distances $d \sim 5r_{\rm vir}$. We have
explicitly checked that this is not due to the averaging over random orbital
phases of individual subhalos; using the apocenter distances instead of the
instantaneous position leads to very similar results. On the other hand, the
tidal field of the host seems to imprint a significant radial alignment on the
subhalos (see the bottom panel of Fig.~\ref{fig:shape-vs-d}), which tend to
orient their major axis radially to the center of the host, albeit with a
large scatter.  The signal is stronger close to the center of the main host
halos and decreases steadily until it disappears at $d \gtrsim 2 r_{\rm vir}$,
where the distribution is consistent with random \citep{Pereira2008,
  Pereira2009}. Interestingly, inner (at $r_{\rm max}$) and $r_{95}$ contours
are well aligned within each object, as shown in the bottom right panel.

Notice that, although our findings suggest only small differences between the
shapes of subhalos and field halos, the evolution of {\it single} objects
under the effects of tidal disruption can differ significantly from the
behavior of the {\it population} as a whole (Barber et al., in preparation.)
In general, the analysis of a population of subhalos such as that shown in
Figs~\ref{fig:shape-vs-vmax}--\ref{fig:shape-vs-d} will be dominated, in
number, by objects with recent infall times (and therefore not largely exposed
to tidal effects), minimizing the differences between subhalos and field halos
in good agreement with our results.

\section{Application to the Modeling of Local Group Satellites}
\label{sec:shape-mw}

As discussed in the Introduction, the Local Group satellite galaxies are
expected to inhabit dark matter subhalos comparable to those studied in the
previous section. Since the contribution of the baryons to the gravitational
potential of these systems is thought to be sub-dominant, the shape, dynamics
and orbital structure of their host dark halos may be compared in a reasonably
direct way to those of a suitable subset of the subhalos in the Aquarius
simulations.

To select subhalos likely to host luminous satellites comparable to local
dwarfs we use the semi-analytical model of \cite{Starkenburg2012}. The
semi-analytic model includes physical prescriptions for the treatment of
relevant processes such as radiative cooling, chemical enrichment, star
formation, supernova feedback, etc. The parameters in the model are tuned to
simultaneously reproduce the luminosity function and spatial clustering of
bright galaxies as well as the properties of satellites in the Local Group
\citep{DeLucia2007, DeLucia2008, Li2010}.

\begin{table}
  \begin{center} \newcolumntype{R}{>{\raggedleft\arraybackslash}X} %
    \begin{tabularx}{0.4\textwidth}{lRRR} \toprule[1.4pt] Parameter & Median &
$-1\sigma$ & $+1\sigma$ \\ \midrule[1.4pt] 
$                                  \alpha$ &       0.27 &       0.07 &       0.10 \\ 
$                    r_{-2}\; ({\rm kpc})$ &       1.21 &       0.42 &       0.58 \\ 
$             v_{-2}\; ({\rm km\;s^{-1}})$ &       9.06 &       1.87 &       1.72 \\ 
$                       c_{\Phi}/a_{\Phi}$ &       0.70 &       0.07 &       0.10 \\ 
$                       b_{\Phi}/a_{\Phi}$ &       0.83 &       0.10 &       0.09 \\ 
$                       r_a\; ({\rm kpc})$ &       1.72 &       1.26 &       2.86 \\ 
$                                    \chi$ &       1.60 &       0.13 &       0.09 \\ 
$                               {\rm ln}A$ &      -2.42 &       0.15 &       0.20 \\ 
\bottomrule[1.4pt]
    \end{tabularx}
  \end{center}
  \caption{Best-fitting values for the profiles shown in
Figs.~\ref{fig:shape-vs-r} and \ref{fig:shape-vs-r-2}. See text for details.}
  \label{tab:fits}
\end{table} 

\subsection{The shapes of subhalos hosting luminous satellites}

\begin{figure}
  \includegraphics[width=0.48\textwidth]{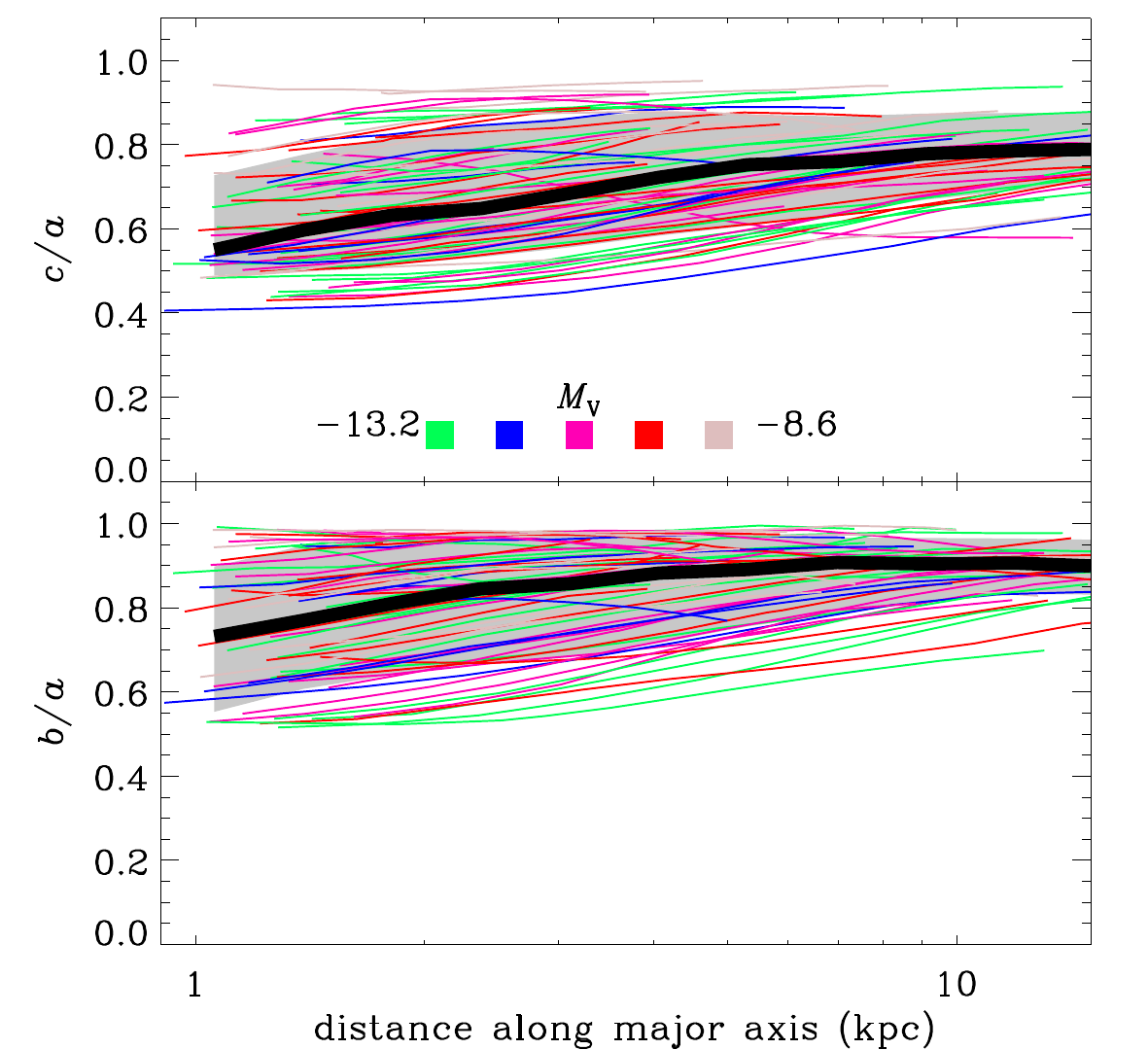}
  \caption{Shape as a function of distance along the major axis, $a$, for
    subhalos hosting luminous satellites. The median profile, and
    $1\sigma$-equivalent percentiles of this sample are given, respectively,
    by the black solid line and the gray shaded area.}
  \label{fig:shape-vs-r}
\end{figure}

Fig.~\ref{fig:shape-vs-r} shows the axis ratios as a function of distance
along the major-axis $r$ for our sample of subhalos. This consists of subhalos
within the virial radius of their hosts at $z=0$ and that resemble the
\emph{classical satellites} of the Milky Way in their luminosity, i.e. their
$V$-band absolute magnitudes are in the range $-13.2 \leq M_V \leq -8.6$. Each
curve is plotted from the convergence radius out to the $r_{95}$ contour, and
the color scale indicates the luminosity assigned by the semi-analytic model
to the satellites.

\begin{figure}
   \begin{center}
     \includegraphics[width=0.48\textwidth]{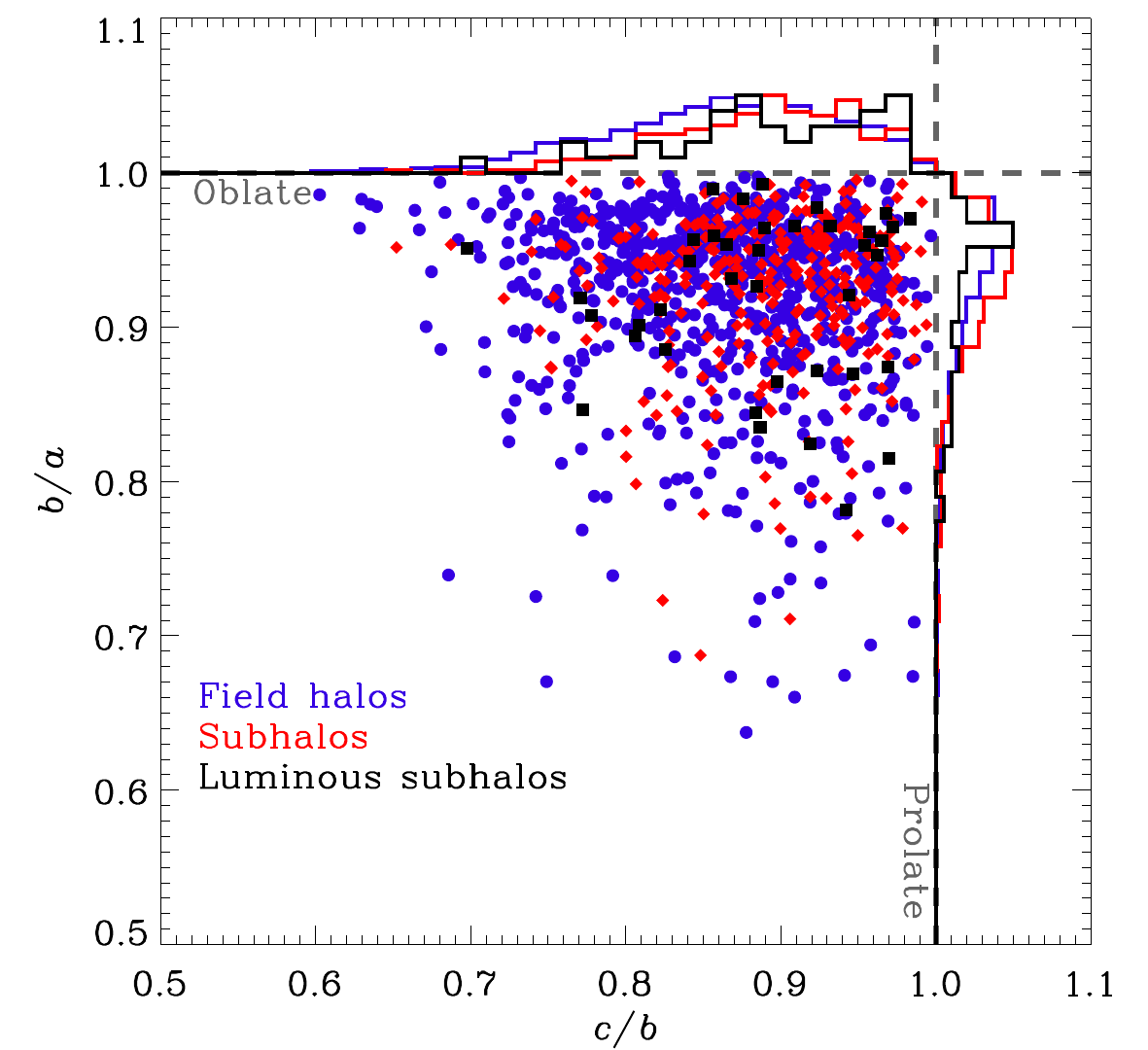}
   \end{center}
   \caption{$b/a$ versus $c/b$ axis ratios measured at $r = 1$ kpc for
     individual (converged) objects. Field halos are shown with blue circles,
     subhalos with red diamonds and luminous subhalos with black squares. The
     histograms show that close to the center subhalos may be approximated by
     oblate axisymmetric objects.}
   \label{fig:p-vs-q}
 \end{figure}

Fig.~\ref{fig:shape-vs-r} shows that the dwarf galaxies in the model are
surrounded by subhalos that are triaxial, with axis ratios $b/a$ and $c/a$
typically increasing from the inner regions to the $r_{95}$ radius. The
scatter from object to object is large, but the overall trend with radius is
similar for all subhalos. The median profile, and $1\sigma$-equivalent
percentiles of the sample are given, respectively, by the black solid line and
the gray shaded area. These dark matter subhalos have on average $c/a \sim
0.60$ and $b/a\sim0.75$ at a radius of $\sim 1$~kpc, and turn more spherical
close to the $r_{95}$ radius, where $c/a\sim 0.8$ and $b/a \sim 0.9$.
Individual inner shapes of halos/subhalos can be clearly seen in
Fig.~\ref{fig:p-vs-q}, where we show a scatter plot of $b/a$ and $c/a$ ratios
measured at $r\sim 1$~kpc.  Only converged objects have been
included. Different symbols are used for different samples: blue circles for
field halos, red diamonds for subhalos and black circles for luminous
subhalos. The fact that subhalos and field halos are well mixed in this plane
confirms the lack of any significant trend between shape and distance to the
main halo, in agreement with Fig.~\ref{fig:shape-vs-d}.

We may use Poisson's equation to derive a relation between the $b/a$ and $c/a$
of the density (which our method measures), and those of the potential
$b_\Phi/a_\Phi$ and $c_\Phi/a_\Phi$. Following \citet{Vogelsberger2008}, we
introduce a generalized radius:
\begin{equation} \widetilde{r} = \frac{r_a + r}{r_a + r_E}r_E,
\label{eq:r}
\end{equation} where $r^2 = x^2 + y^2 + z^2$ is the Euclidean distance, $r_E^2
= (x/a_{\Phi})^2 + (y/b_{\Phi})^2 + (z/c_{\Phi})^2$ is the \emph{ellipsoidal}
radius and $r_a$ a characteristic scale. With this definition, $\widetilde{r}
\approx r_E$ for $r \ll r_a$ and $\widetilde{r} \approx r$ for $r \gg r_a$.
Assuming that the potential at any point is
\begin{equation} \Phi(x,y,z) = \widetilde{\Phi}(\widetilde{r}),
\label{eq:phi}
\end{equation} where $ \widetilde{\Phi}$ is the spherically symmetric
potential associated with the Einasto profile \citep{Einasto1965}, we find for
our sample a median $b_\Phi/a_\Phi = 0.83$ and $c_\Phi/a_\Phi = 0.70$. The
median and $\pm 1\sigma$ error of the parameter fits for the density and axis
ratios obtained in this way are given in Table~\ref{tab:fits}.

\begin{figure}
  \includegraphics[width=0.48\textwidth]{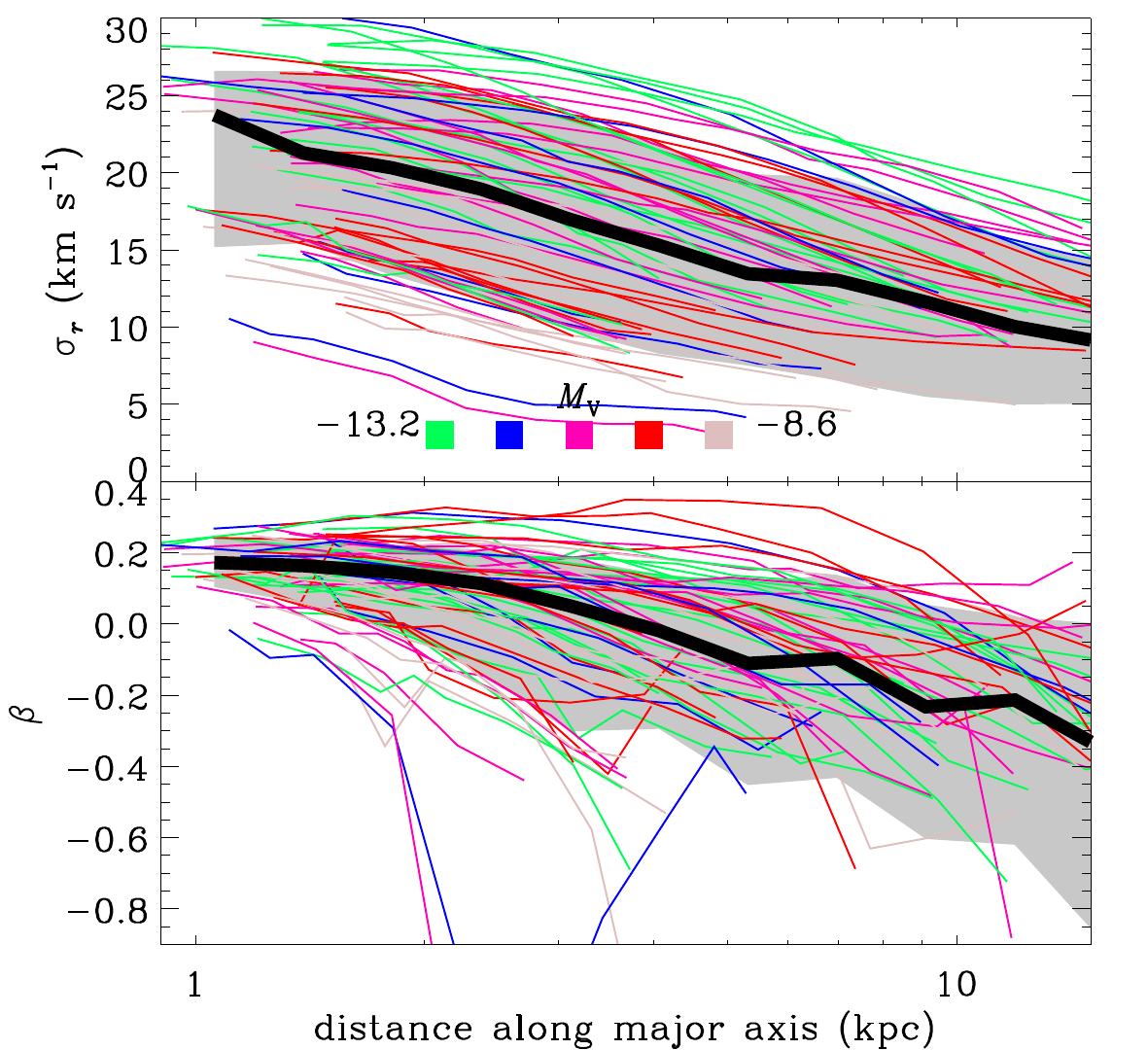}
  \caption{Radial velocity dispersion and anisotropy as a function of the
    distance along the major axis, $r$, for the subhalos hosting luminous
    satellites. The thick solid black line represents the median behavior.}
  \label{fig:shape-vs-r-2}
\end{figure}

\subsection{The internal kinematics of subhalos hosting luminous satellites}
\subsubsection{Behavior along the major axis}

Fig.~\ref{fig:shape-vs-r-2} shows the radial velocity dispersion $\sigma_r$
(top) and the orbital anisotropy $\beta$ (bottom), both as a function of
distance along the major axis. These quantities are computed in ellipsoidal
coordinates that follow the axis ratios of the mass density at each radius.
In practice, we calculate the component of the velocity in the direction
tangential to a given ellipsoid, $\sigma_T$, and the radial component
$\sigma_r$ is derived by subtraction in quadrature of $\sigma_T$ from the
total velocity dispersion.  Therefore this $\sigma_r$ corresponds very closely
to the velocity dispersion along the spherical radial direction as it is
computed along the ellipsoid's major axis.

As in the previous figure, individual objects are shown with thin lines
color-coded according to their $V$-band absolute magnitude. The median trend
of the sample is indicated by a black solid curve, $1\sigma$-equivalent
percentiles in gray shading.  The spread in $\sigma_r$ seen in the top panel
of Fig.~\ref{fig:shape-vs-r-2} is due to the difference in mass of the
subhalos, which span a range $m=1.6\times 10^8- 5.8\times 10^9\; {\rm
  M}_\odot$.

A declining velocity dispersion profile as seen in the top panel of
Fig.~\ref{fig:shape-vs-r-2} is generally expected for halos in CDM because of
the relation between the density and pseudo phase-space density, namely
$Q=\rho/\sigma_r^3$ \citep{Taylor2001}:
\begin{equation} \frac{\varrho}{\widetilde{\sigma}_r^3} = A x^{-\chi},
  \label{eq:q}
\end{equation} with $A$ a normalization constant, $\varrho \equiv
\rho/\rho_{-2}$, $\widetilde{\sigma}_r \equiv \sigma_r / v_{-2}$ and $x \equiv
r/r_{-2}$, where $\rho_{-2}$, $v_{-2}$ and $r_{-2}$ are the characteristic
density, velocity and radius for the Einasto density profile, 
respectively\footnote{See Appendix \ref{sec:appendix-a}, where we show that a
  power-law fit is a reasonable description of the pseudo-phase-space of
  subhalos just like it is for field halos.}. When we fit the $Q$-profiles
individually for each subhalo, we obtain median values of $\ln A=-2.42$ and
$\chi=1.60$, the latter indicating a slightly shallower fall off than for
field halos. 

In the bottom panel of Fig.~\ref{fig:shape-vs-r-2} we plot the ellipsoidal
velocity anisotropy $\beta$ profiles. Here $\beta = 1 -
\sigma_r^2/\sigma_T^2$, where we calculate the component of the velocity
tangential and radial to a given ellipsoid as explained above. The velocity
anisotropy profiles of dark matter subhalos tend to decline with radius. In
the inner regions the motions are slightly radially-biased ($\beta \sim 0.2$
at $r \sim 1$ kpc), while the ellipsoid becomes increasingly tangential
($\beta < 0$) at larger radii. This behavior is markedly different from the
radially-biased ellipsoids found in isolated $\Lambda$CDM halos, particularly
in the outskirts \citep{Cole1996,Taylor2001, Wojtak2005,Ludlow2010}.  This
difference is a result of tidal forces, which preferentially remove particles
with large apocenters on radial orbits. Fig.~\ref{fig:shape-vs-r-2} also shows
that subhalos rarely have a constant $\beta$ profile.

We may derive an expression for $\beta$ using the spherical Jeans Equation
which relates the density, anisotropy and radial velocity dispersion of a
system \citep{Binney_Tremaine2008}. In the case of an Einasto profile with
logarithmic slope $\sim -r^{\alpha}$ we find:
\begin{equation} 3\beta (r) = - \frac{6\pi \mu}{x \widetilde{\sigma}_r^2} +
5x^{\alpha} - \chi,
  \label{eq:beta}
\end{equation} where $\widetilde{\sigma}_r = \sigma_r / v_{-2}$ as before and
$\mu=\mu(x,\alpha)$ is given by Eq.~\eqref{eqn:apdx-mass-einasto-dim} (see
Appendix \ref{sec:appendix-a} for a more detailed derivation of this
expression).  The velocity anisotropy is therefore dependent on the
logarithmic slopes of the mass density and of the pseudo phase-space density,
$\alpha$ and $\chi$, respectively.

This expression provides a reasonable fit out to $r \sim 2.5 r_{-2}$. Whereas
the limiting behavior of the anisotropy in the inner regions is similar for
all subhalos, beyond a radius of $\sim 1$~kpc large variations are seen from
object-to-object. These variations are still accounted for by
Eq.~\eqref{eq:beta} when each $\beta$ profile is fitted individually. We find
that the exact shape of the anisotropy profile depends most strongly on
$\alpha$, while $\chi$ determines where the velocity ellipsoid becomes
tangential at large radii. On the other hand, variations in $\ln A$ have a
very minor effect.

\subsubsection{Axisymmetric description of the internal kinematics}

Fig.~\ref{fig:shape-vs-r} shows that subhalos are not significantly triaxial
having in average $b/a\gtrsim 0.8$, therefore, an axisymmetric approach may be
sufficient to describe their internal kinematics.  Moreover, an oblate
approximation seems favoured by the overall distribution of inner shapes shown
in Fig.~\ref{fig:p-vs-q}, albeit with a large scatter.

\begin{figure}
  \includegraphics[width=0.48\textwidth]{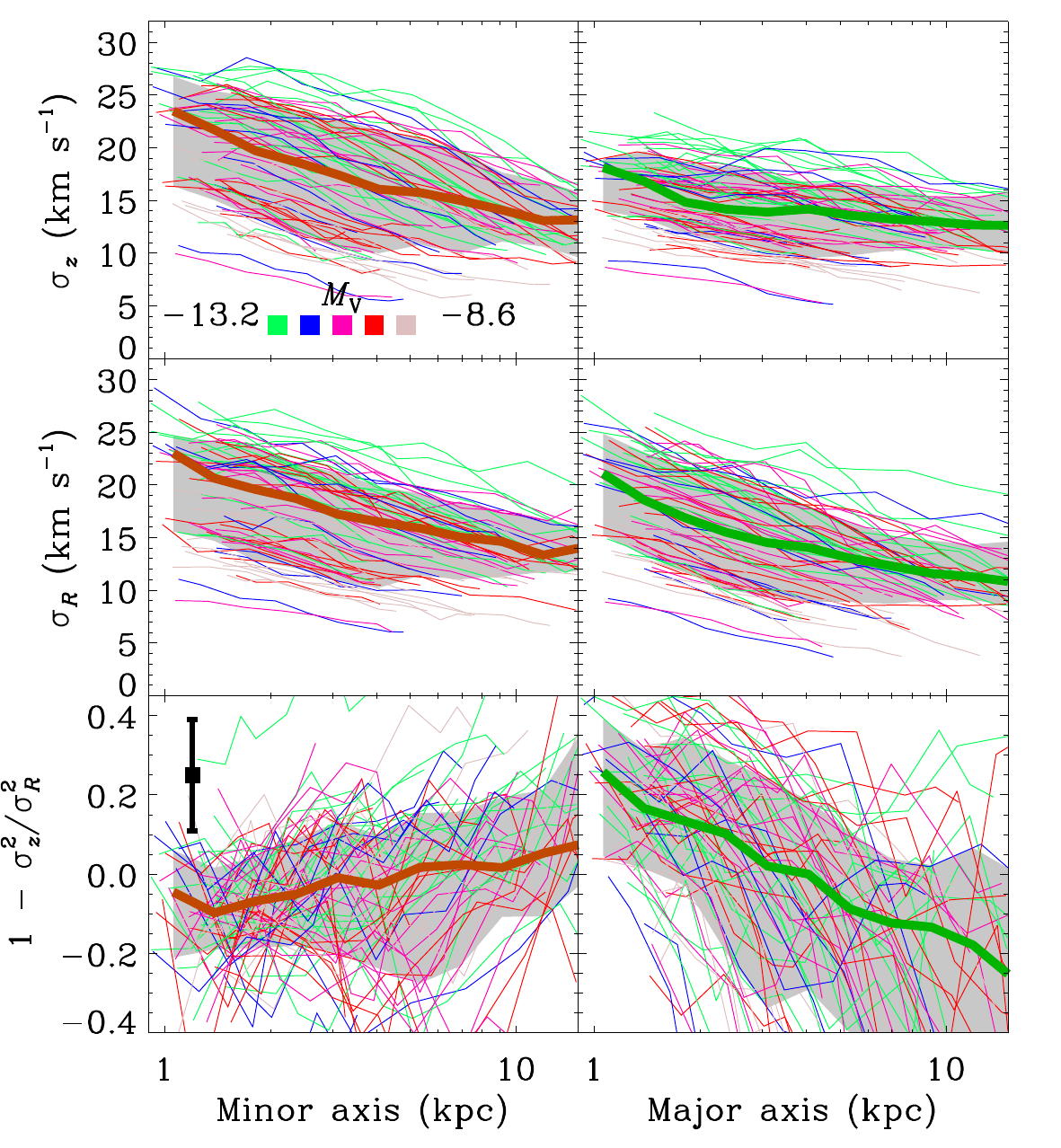}
  \caption{Velocity dispersion as a function of distance along the minor
    (left) and major axis (right) for subhalos hosting luminous satellites.
    In each case the system is rotated such that the minor axis lies along the
    $z$ direction.  The cylindrical velocity dispersions $\sigma_z^2$ and
    $\sigma_R^2$ are estimated in spheres containing 400 particles.}
  \label{fig:sigma-vs-r}
\end{figure}

To this end, in Fig.~\ref{fig:sigma-vs-r} we explore the velocity structure in
the cylindrical radial and vertical directions. Each object is rotated such
that the minor axis coincides with the $z$-direction. The velocity dispersion
components along the vertical ($z$) and radial ($R$) axes provide information
on how dynamically hot a system is in both directions. We compute this along
the two preferential axis, minor (left) and major (right), using at each
radius a sphere that contains 400 particles. The velocity dispersion
$\sigma_z$ and $\sigma_R$ are then computed within these volumes and displayed
as a function of distance along the axes. As before, individual halos are
shown in thin blue lines, the median values with solid thick lines.

The velocity dispersion profiles show a steady decline with radius, and as
expected with a trend similar to the radial velocity dispersion shown in
Fig.~\ref{fig:shape-vs-r}. As in that case, there is scatter from
object-to-object mostly due to the range in subhalo mass, with the most
luminous objects (light green in Fig.~\ref{fig:sigma-vs-r}) having larger
velocity dispersion than the fainter ones (red). The anisotropy $\beta_z$ is
defined as $\beta_z=1-\sigma_z^2/\sigma_R^2$ and is shown in the bottom panels
of Fig.~\ref{fig:sigma-vs-r}. Interestingly, along the minor axis we find
$\sigma_z \sim \sigma_R$, in agreement with the assumptions by
\citet{Hayashi2012} when modeling stars in the dSph. However, this is not true
for the velocity dispersions along the major axis as shown in the right panel,
where we find positive values of $\beta_z$ close to the center and $\beta_z
<0$ at the $r_{95}$ radius. This trend, although systematic, is quite weak,
with large scatter amongst the individual objects.  It is important to realize
that our measurements are subject to significant noise as we are now
restricted to consider a small region around a given radius (i.e. we do not
average over entire ellipsoidal shells as before and so each sphere contains
fewer particles). The typical uncertainty for the individual curves is shown
by the black vertical error bar, computed as the dispersion obtained from
drawing 100 samples with replacement in each bin.

The fact that the velocity dispersions $\sigma_R$ and $\sigma_z$ are not equal
implies that the subhalos' distribution functions are a function of a third
integral. Although we have shown this to be the case for dark matter
satellites, it could also be true for the stars embedded in these
systems. Therefore dynamical models of dSph may need to take this into account
\citep{Battaglia2013}, since neglecting this fact can lead to unrealistic
estimates of the shapes of the host dark matter halos
\citep[see][]{Hayashi2012}.

\begin{figure}
  \includegraphics[width=0.48\textwidth]{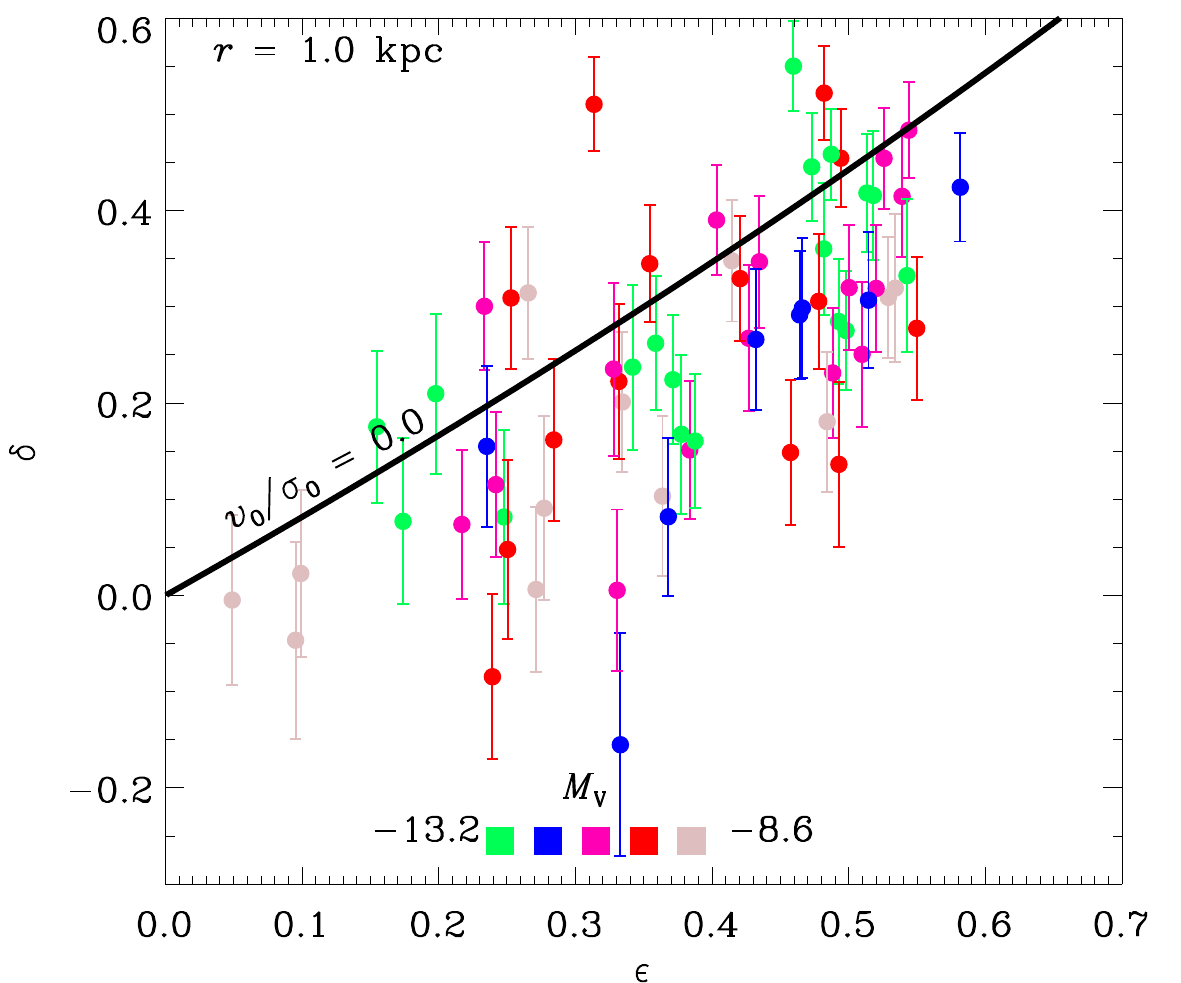}
  \caption{Relation between the anisotropy $\delta = 1 - \sigma_{z}^2 /
    \sigma_x^2$ and the ellipticity of the halos $\epsilon=1-c/a$ measured at
    a distance $x= 1$ kpc along the major axis. The solid black line shows the
    prediction of the virial tensor theorem when applied to axysimmetric
    systems supported by random motion \citep{Binney2005}.}
  \label{fig:shape-velocity}
\end{figure}

The shapes of the subhalos are consistent with dynamical support by the
velocity ellipsoid, as shown by Fig.~\ref{fig:shape-velocity}.  The vertical
axis shows the local anisotropy $\delta = 1-\sigma_{z}^2/\sigma_{x}^2$, where
$x$ is again the direction of the major axis and $z$ points along the minor
axis. These quantities are calculated in a sphere with 400 particles located
at $x=1$ kpc and the individual points in this figure correspond to the
different subhalos.  In the axisymmetric case, the virial theorem in tensor
form \citep{Binney2005} gives:
\begin{equation} \frac{v_0^2}{\sigma_0^2} = 2(1-\delta) \frac{W_{xx}}{W_{zz}}
-2,
\label{eq:binney}
\end{equation}

\noindent where $W_{ij}$ are the components of the potential-energy tensor
\citep{Binney_Tremaine2008}. For ellipsoidal systems $W_{xx}/W_{zz}$ is a
function of the ellipticity $\epsilon=1-c/a$ and is independent of the radial
density profile \citep{Roberts1962}. $v_0$ is the streaming velocity along the
$y$-axis and $\sigma_0$ is the velocity dispersion in the $x$ direction. The
solid line in Fig. 9 indicates the prediction from Eq.~\eqref{eq:binney} in
the case of a dispersion supported system with $v_0/\sigma_0 = 0$.  This
prediction provides a reasonable representation of the simulated objects that
agrees well with the very little rotation that we find: halos and subhalos
show an average $\langle v_0/\sigma_0 \rangle = 0.08$ and more than $90\%$ of
the sample has $v_0/\sigma_0 < 0.14$ in their inner regions. The scatter,
however, is large and cannot be explained solely on the basis of rotation of
subhalos at small radii.  Further factors such as departure from axisymmetry
or the lack of dynamical equilibrium generated by tides may also contribute to
the scatter seen in Fig.~\ref{fig:shape-velocity}.

\section{Conclusions}
\label{sec:conclusions}

We have used the Aquarius simulations to study the shapes of field and
satellite dark matter halos with emphasis on the mass range expected for the
hosts of the dwarf galaxies in the Local Group. We have used an iterative
method based on the normalized inertia tensor to characterize the principal
axis lengths $a \ge b \ge c$ of halos and subhalos as a function of radius. In
particular, we have explored in detail halo shapes measured in the inner
regions (radius of maximum circular velocity $r_{\rm max}$) and in the
outskirts or \emph{$r_{95}$ contour}. Although stars are more centrally
concentrated than the dark matter, our resolution allows us to characterize
halo shapes at radii as small as $r \sim 1 \; \rm kpc$, starting to probe the
regime traced by the outer stars in dwarf galaxies.

Through a comparison of objects in common between the different resolution
levels of the Aquarius simulations, we have noticed that simple number of
particles cuts do not guarantee convergence in the measured halo shapes,
especially in the inner regions. We find that instead the \emph{convergence
radius} $r_{\rm conv}$ (defined as the threshold $\kappa = 7$ in the ratio
between the local relaxation time and the dynamical time at the virial radius
\citep{Power2003}) provides a good estimate of the radius where the axis
ratios are robustly determined (with an error $< 8$\%).

We have found that the typical axis ratios of isolated halos in the Aquarius
simulations decrease with increasing mass, or equivalently maximum halo
circular velocity $V_{\rm max}$, i.e.\ low mass objects tend to be more
spherical than Milky Way-like objects.  These trends are well approximated by
a relation between the axis ratio measured at the $r_{95}$ radius
$c/a|_{r_{95}}$ and $V_{\rm max}$, i.e.  $c/a|_{r_{95}} \sim -0.021 \log
V_{\rm max}$, while this relation is slightly steeper if the axis ratio is
measured at $r_{\rm max}$ contour in which case $c/a|_{r_{\rm max}} \sim
-0.032 \log V_{\rm max}$.  The differences in the shapes of field versus
satellite halos are small and within the intrinsic scatter of the samples.
Nonetheless, at a fixed $V_{\rm max}$, subhalos tend to have larger axis
ratios than isolated objects in the field.

The similarity between subhalos and field objects is also apparent in the lack
of significant trends in the axis ratios with distance to the main host halo,
$d$. We find, however, that the alignment of the ellipsoids varies with $d$:
dark matter halos at close distances from the host center tend to be oriented
preferentially with their major axis pointing radially. The signal disappears
only for $d \gtrsim 2.5$ $r_{\rm vir}$, where the orientations are consistent
with random.

We have also focused on the properties of subhalos likely to host analogs of
the \emph{classical satellites} of the Milky Way ($-13.2 \leq M_V \leq -8.6$),
according to the semianalytic model of galaxy formation run on the Aquarius
suite by \citet{Starkenburg2012}. Our analysis indicates that these galaxies
are hosted by mildly triaxial dark matter objects with minor-to-major axis
ratios $c/a\approx 0.60$ and intermediate-to-major $b/a\approx 0.75$ in the
first kiloparsec with a clear trend towards becoming axisymmetric in the
outskirts. Their internal orbital structure shows evidence of being affected
by tidal forces from their hosts (i.e. the main Aquarius halos), since the
velocity anisotropy becomes tangential with radius, in clear contrast to what
is found for isolated systems. We have also found that this orbital structure
may be modeled in the axisymmetric context, where the velocity anisotropy
$\beta_z \sim 0 $ along the minor axis, and declines with distance along the
major axis. These results may be used to motivate more realistic models of the
subhalos hosting satellite galaxies like those observed around the Milky Way.

\section*{Acknowledgements}
\label{acknowledgements}

We thank Else Starkenburg for access to the semi-analytical catalog of the
Aquarius simulations. AH gratefully acknowledges financial support from the
European Research Council under ERC-Starting Grant GALACTICA-240271.  The
authors thank the hospitality of the Kavli Institute for Theoretical Physics,
Santa Barbara during the program ``First Galaxies and Faint Dwarfs: Clues to
the Small Scale Structure of Cold Dark Matter'', where part of this work was
completed under the support of the National Science Foundation Grant No. NSF
PHY11-25915. This work was supported in part by the National Science
Foundation under Grant No. PHYS-1066293 and the hospitality of the Aspen
Center for Physics.

\bsp
\label{lastpage}

\bibliographystyle{mn2e} \bibliography{refs}

\appendix
\section{Einasto Profiles and The spherical Jeans equation}
\label{sec:appendix-a}

Following the method of \citet{Vera2012} we fit a Einasto profile to the
circular velocity profile of each individual subhalo of our sample.  For each
object we take 20 bins equally spaced in logarithmic space between $r_{\rm
conv}<r<0.9r_{95}$.  We compute the cumulative circular velocity
profile and fit an Einasto model by minimizing the merit function $E =
\sum_{i=1}^{N_{\rm bins}} (\ln v_c^2(r_i) - \ln v_{c,i}^2)^2/N_{\rm bins}$
against the free parameters $r_{-2}$, $\rho_{-2}$ and $\alpha$.  Here,
$v_{c,i}$ is the circular velocity corresponding to an Einasto profile with a
scale radius $r_{-2}$ (the radius at which the density profile has a slope
$-2$), a characteristic density at $r_{-2}$ equal to $\rho_{-2}$ and a shape
parameter that controls the overall slope of the profile, $\alpha$.  We
deliberately chose the circular velocity profile over the more widely used
density profile which is more sensitive to shot noise in each bin
\citep[see][for more details]{Vera2012}.

The density profile can be written as,
\begin{equation}\label{eqn:apdx-rho-einasto} \rho (r) =
\rho_{-2}\varrho(r/r_{-2}),
\end{equation}
  
\noindent where
\begin{equation}\label{eqn:apdx-rho-einasto-dim} {\rm ln }\varrho(x) =
-\frac{2}{\alpha}(x^{\alpha} -1).
\end{equation}

\noindent $\varrho$ is therefore a dimensionless function of the dimensionless
variable $x = r/r_{-2}$. In this spirit, it is possible to define a set of
scaling factors in which we can express the dynamics of the system, namely, a
characteristic mass $m_{-2} \equiv r_{-2}^3 \rho_{-2}$ and characteristic
velocity $v_{-2}^2 \equiv Gr_{-2}^2\rho_{-2}$ [note that $v_{-2}$ is not
$v_{\rm circ}(r_{-2})$]. The enclosed mass within a radius $r$ is therefore
$m(r) = 4\pi m_{-2} \mu(r/r_{-2})$, where
\begin{eqnarray}\label{eqn:apdx-mass-einasto-dim} \mu(x) &=& \int_0^x {\rm d}x\; x^2
\varrho(x) \nonumber \\ &=& \frac{1}{\alpha}\exp\left( \frac{3\ln\alpha + 2 -
\ln 8}{\alpha}\right) \gamma \left(\frac{3}{\alpha}, \frac{2x^\alpha}{\alpha}
\right),
\end{eqnarray}

\noindent with $\gamma$ the lower incomplete gamma function.  In a similar
fashion we can define a dimensionless version of the radial velocity
dispersion $\widetilde{\sigma}_r(x) \equiv \sigma_r(r_{-2}x)/{v_{-2}}$.

\begin{figure}
  \includegraphics[width=0.48\textwidth]{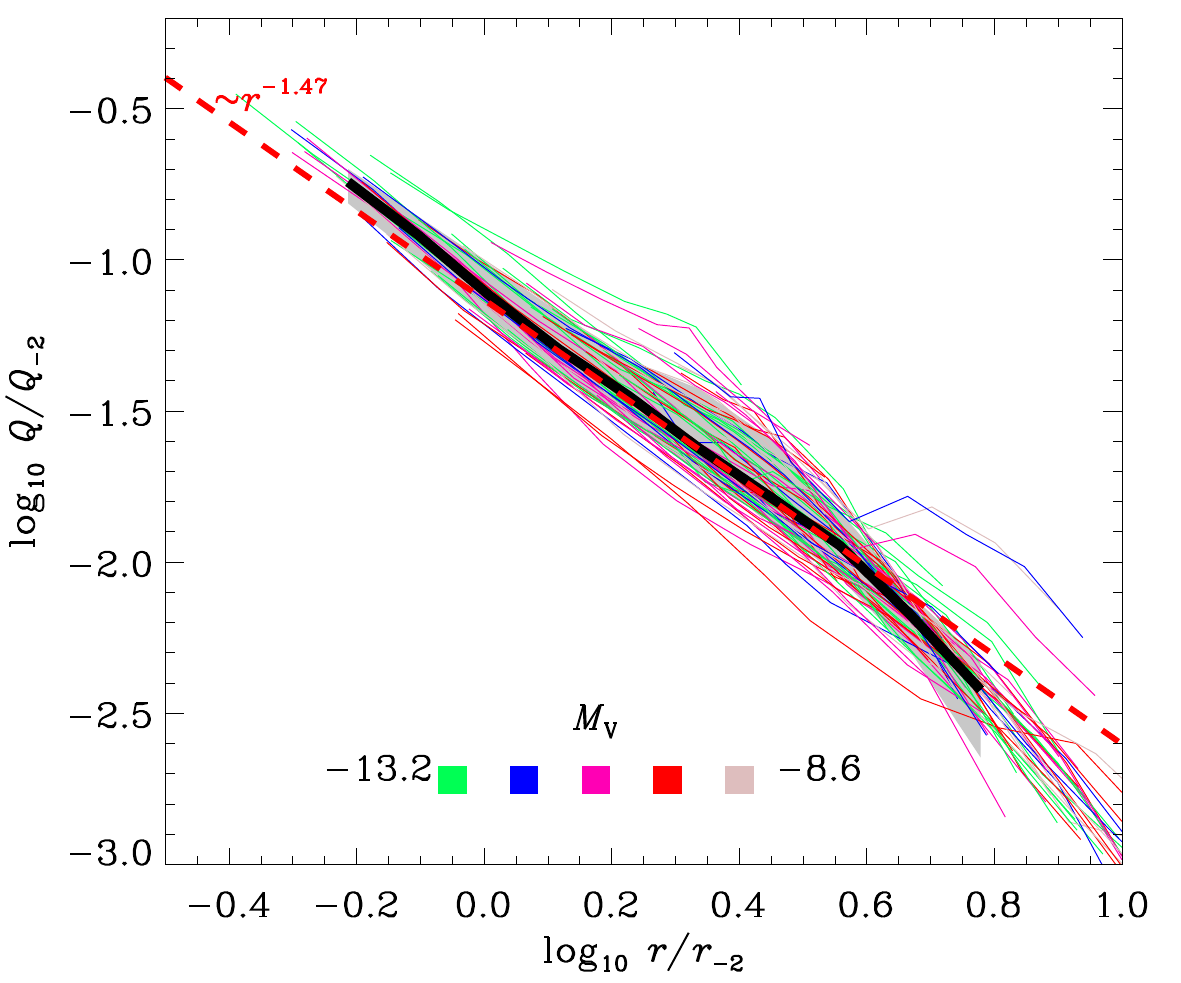}
  \caption{Pseudo-phase-space density profile $Q/Q_{-2} =
    \varrho/\widetilde{\sigma}_r^3$ for the subhalos hosting luminous dwarfs
    in our sample. The black line shows the mean profile of all subhalos and
    the red curve indicates the best power law fit $Q \sim r^{-1.47}$.}
  \label{fig:ppsd}
\end{figure}

It has been previously reported in the literature that the pseudo phase-space
density profile of isolated dark matter halos can be well modeled by a single
power law $Q \equiv \rho/\sigma_r^2\sim r^{-\chi}$, $\chi>0$
\citep{Taylor2001,Dehnen2005, Navarro2010, Ludlow2011}. Fig.~\ref{fig:ppsd}
shows $Q$ measured for our sample of subhalos hosting luminous satellites. We
have found that the slope $\chi$ is slightly shallower than for objects in the
field ($\chi\sim1.6$ versus $\chi \sim 1.8$ for isolated halos). The
best-fitting values of the parameters, and their variance, are given in
Table~\ref{tab:fits}.

Notice that the pseudo-phase-space profiles start to deviate from a power-law
in the outer regions, likely induced by ongoing tidal stripping. This
typically occurs for $\log r/r_{-2} \geq 0.6$, which is roughly the same scale
at which our fit for the radial velocity dispersion $\sigma_r$ deviates from
the mean subhalo trends shown in Fig.~\ref{fig:shape-vs-r-2}.

Using the dimensionless quantities introduced before, we can now write the
spherical Jeans equation as:
\begin{eqnarray}\label{jeans-model} && \frac{d}{dx}(\varrho
\widetilde{\sigma}_r^2) + 2\frac{\beta }{x}(\varrho \widetilde{\sigma}_r^2) =
-\frac{4\pi \varrho \mu}{x^2} \nonumber \\ &\Rightarrow &
\frac{d\ln\varrho}{d\ln x} + 2\frac{d\ln\widetilde{\sigma}_r}{d\ln x} + 2\beta
= -\frac{4\pi\mu}{x \widetilde{\sigma}_r^2}.
\end{eqnarray}

\noindent We can use Eq.~\eqref{eqn:apdx-rho-einasto-dim} to further reduce
this expression:
\begin{equation}\label{eq:model-ein} 3\beta (x) =
-\frac{6\pi\mu}{x\widetilde{\sigma}_r^2} + 5x^{\alpha} - \chi.
\end{equation}

The limiting value of this expression at small radii can be obtained from a
Taylor expansion around zero. For $x\ll 1$ we may use that $\lim_{x \to 0}
\gamma(s,x)/x^s=1/s$ \citep{Abramowitz1972}. Finally
\begin{eqnarray*} \ln \varrho(x) &\approx& \frac{2}{\alpha},\\ \mu(x)
&\approx& \frac{1}{3}x^3{\rm e}^{2/\alpha},\\ \widetilde{\sigma}_r(x)
&\approx& A^{-1/3}{\rm e}^{2/3\alpha} x^{\chi/3}, \qquad x\ll 1,
\end{eqnarray*}
\noindent which leads to

\begin{equation} 3\beta(x) \approx -2\pi A^{2/3}{\rm
e}^{2/3\alpha}x^{2(1-\chi/3)} + 5x^{\alpha} - \chi, \qquad x\ll 1.
\end{equation}

\end{document}